\documentstyle[preprint,aps,floats,epsfig]{revtex}
\tightenlines

\begin{document}
\title{Charm meson production from meson-nucleon scattering}
\bigskip
\author{W. Liu and C. M. Ko}
\address{Cyclotron Institute and Physics Department, Texas A\&M University,
College Station, Texas 77843-3366}
\maketitle

\begin{abstract}
Using an effective hadronic Lagrangian with physical hadron masses
and coupling constants determined either empirically or from 
SU(4) flavor symmetry, we study the production cross sections of 
charm mesons from pion and rho meson interactions with nucleons. 
With a cutoff parameter of 1 GeV at interaction vertices as usually 
used in studying the cross sections for $J/\psi$ absorption and 
charm meson scattering by hadrons, we find that the cross sections 
for charm meson production have values of a few tenth of mb and 
are dominated by the $s$ channel nucleon pole diagram.  
Relevance of these reactions to charm meson production in 
relativistic heavy ion collisions is discussed.

\end{abstract}
 
\section{introduction}

Because of their large masses, open charm mesons are expected to 
be mostly produced in the initial preequilibrium stage of
relativistic heavy ion collisions. They have thus been suggested as
possible probes of the initial dynamics in these collisions.
Previous studies have been concentrated on the production of charm quarks 
from the preequilibrium partonic matter \cite{ling,linv}. In these studies,
it has been found that charm quark production is sensitive to not only
the rapidity and space correlations of initial minijet partons but
also their energy loss in the dense partonic matter.  For
charm meson production from nonpartonic matter, the only study is 
the one \cite{hcharm} based on the Hadron-String Dynamics (HSD) \cite{hsd} 
using hadronic cross sections obtained from the Quark-Gluon String Model 
(QGSM) \cite{qgsm}. Allowing scatterings between the leading quark and 
diquark in a baryonic string with the quark and antiquark in a mesonic 
string and taking their cross sections to be the same as in meson-baryon 
scatterings, this study shows that charm production is appreciable 
even with a small cross section of a few $\mu$b as predicted by the QGSM. 
The factor of two enhancement obtained in this study for charm
mesons over that produced from the primary nucleon-nucleon 
collisions offers a possible explanation for the observed enhancement 
of intermediate mass dileptons seen in heavy ion collisions at SPS 
\cite{abreu}.  

The QGSM model treats charm meson production from pion-nucleon scattering
as a process involving the exchange of the vector charm meson Regge 
trajectory in $t$-channel.  Contributions from the $s$ and $u$ channels 
are neglected. Although the $u$ channel is expected to be small as it 
involves nonplanar diagrams, which are known to be negligible in the large 
$N_c$ limit, the $s$ channel contribution may not be small
because of the planarity of associated diagrams.  To study the relative 
importance of the $s$, $t$, and $u$ channel contributions to charm meson 
production in pion-nucleon scattering, we use an effective
hadronic Lagrangian based on the flavor SU(4) symmetry but with empirical 
hadron masses. This Lagrangian has recently been used to study
the cross sections for both $J/\psi$ absorption 
\cite{matinyan,haglin,lin,oh,liu} 
and charm meson scattering \cite{cscatt} by hadrons.
We find that the magnitude of the cross section for charm meson 
production from pion-nucleon scattering depends sensitively on the
value of the cutoff parameter at interaction vertices. Using a cutoff 
parameter of 1 GeV as used previously in studying $J/\psi$ absorption
and charm meson scattering, we find that the $t$ channel process 
involving vector charm meson exchange indeed gives a small cross section 
as in QGSM and the $u$ channel contribution is negligible. 
The contribution from the $s$ channel is, however, appreciable,
leading to a few tenth of mb for the production cross section 
of charm meson from pion-nucleon scattering. Furthermore, 
the model allows us to study the cross section for charm production 
from the interaction of nucleons with rho mesons, which are abundant
in the initial stage of the hadronic matter in heavy ion collisions
and also have a lower threshold for charm meson production.
 
This paper is organized as follows. In Section \ref{lagrangian}, we 
introduce the effective interaction Lagrangians needed for studying 
charm meson production from pion-nucleon and rho-nucleon scattering. 
Cross sections for theses processes are then derived in Section 
\ref{cs} and evaluated in Section \ref{results}.  Finally, a summary 
is given in Section \ref{summary}. 

\section{interaction Lagrangians}\label{lagrangian} 

\begin{figure}[ht]
\centerline{\epsfig{file=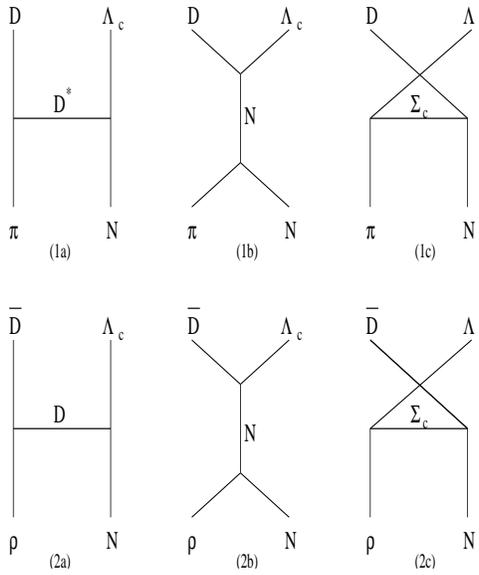,width=2.5in,height=3in,angle=0}}
\vspace{0.5cm}
\caption{Charm meson production from meson-nucleon scattering.}
\label{diagram}
\end{figure}

Possible processes for charm meson production from meson-nucleon 
scattering are $\pi N\rightarrow\bar{D}\Lambda_{c}$ and 
$\rho N\rightarrow \bar{D}\Lambda_{c}$ as shown by the diagrams in 
Fig. \ref{diagram}. For both pion-nucleon and rho-nucleon reactions,
there are $t$ channel charm meson exchange diagrams, $s$ channel
nucleon pole diagrams, and $u$ channel charm baryon pole diagrams.
Cross sections for these 
processes can be evaluated using the same Lagrangian introduced 
in Ref. \cite{cscatt} for studying charm meson scattering by hadrons.
This Lagrangian is based on the gauged SU(4) flavor symmetry but with 
empirical masses. The coupling constants are taken, if possible, from 
empirical information. Otherwise, the SU(4) relations are used to 
relate the unknown coupling constants to the known ones. 

The interaction Lagrangian densities that are relevant to the
present study are given as follows:
\begin{eqnarray} 
{\cal L}_{\pi NN} & = & -\frac{f_{\pi NN}}{m_{\pi}}\bar{N}\gamma_{5}
\gamma^{\mu}\vec{\tau}N\cdot\partial_{\mu}\vec{\pi},\\ 
{\cal L}_{\rho NN} & = & g_{\rho NN}\bar{N}(\gamma^{\mu}\vec{\tau}\cdot
\vec{\rho}_{\mu}+\frac{\kappa_{\rho}}{2m_{N}}\sigma_{\mu\nu}\vec{\tau}
\cdot\partial_{\mu}\vec{\rho}_{\nu})N, \\
{\cal L}_{\pi DD^{*}} & = & ig_{\pi DD^{*}}D^{*\mu}\vec{\tau}\cdot(\bar{D}
\partial_{\mu}\vec{\pi}-\partial_{\mu}\bar{D}\vec{\pi})+{\rm H.c.}, \\
{\cal L}_{\rho DD} & = & ig_{\rho DD}(D\vec{\tau}\partial_{\mu}\bar{D}-
\partial_{\mu}D\vec{\tau}\bar{D})\cdot\vec{\rho}^{\mu},\\  
{\cal L}_{DN\Lambda_{c}} & = &\frac{f_{DN\Lambda_{c}}}{m_D}
(\bar{N}\gamma_{5}\gamma^{\mu}\Lambda_{c}\partial_{\mu}D+\partial_{\mu}\bar{D}
\bar{\Lambda}_{c}\gamma_{5}\gamma^{\mu}N),\\
{\cal L}_{D^{*}N\Lambda_{c}} & = &-g_{D^{*}N\Lambda_{c}}
(\bar{N}\gamma_{\mu}\Lambda_{c}D^{*\mu}+\bar{D}^{*\mu}\bar{\Lambda}_{c} 
\gamma_{\mu}N),\\
{\cal L}_{\pi\Sigma_c\Lambda_c} & = &\frac{f_{\pi \Sigma_c\Lambda_c}}{m_{\pi}}
[\bar{\Lambda}_c\gamma^5\gamma^{\mu}{\rm tr}(\vec{\tau}\cdot\vec
{\Sigma}_c\vec{\tau}\cdot\partial_{\mu}\vec{\pi})
+{\rm tr}(\vec{\tau}\cdot\bar{\vec{\Sigma}}_c\vec{\tau}\cdot
\partial_{\mu}\vec{\pi})\gamma^5\gamma^{\mu}\Lambda_{c}],\\
{\cal L}_{\rho\Sigma_c\Lambda_c} & =& g_{\rho\Sigma_c\Lambda_c}
[\bar{\Lambda}_c\gamma^{\mu}{\rm tr}(\vec{\tau}\cdot\vec{\Sigma}_c
\vec{\tau}\cdot\vec{\rho}_{\mu})
+{\rm tr}(\vec{\tau}\cdot\bar{\vec{\Sigma}}_c\vec{\tau}\cdot
\vec{\rho}_{\mu})\gamma^{\mu}\Lambda_c],\\
{\cal L}_{DN\Sigma_c} & = & \frac{f_{DN\Sigma_c}}{m_D}(\bar{N}\gamma_{5}
\gamma^{\mu}\vec{\tau}\cdot\vec{\Sigma}_c\partial_{\mu}\bar{D}
+\vec{\tau}\cdot\bar{\vec{\Sigma}}_c\gamma_{5}
\gamma^{\mu}N\partial_{\mu}D).        
\end{eqnarray}
In the above, ${\vec\tau}$ are Pauli matrices for isospin, and $\vec{\pi}$ 
and $\vec{\rho}$ denote the pion and rho meson isospin triplet, respectively,
while $D=(D^0,D^+)$ and $D^*=({D^*}^0,{D^*}^+)$ denote the pseudoscalar
and vector charm meson doublets, respectively. 

For coupling constants, we use the empirical values for 
$f_{\pi NN}/m_{\pi}=7.18$ GeV$^{-1}$\cite{pnn}, $g_{\rho NN}=3.25$, 
and $\kappa_\rho=6.1$ \cite{rnn}. From the recently measured width 
$\Gamma_D^*\sim 96$ keV of $D^*$ \cite{ahmed}, we obtain 
the coupling constant $g_{\pi DD^*}=5.56$, which is slightly
larger than the one used in Refs. \cite{matinyan,haglin,lin,oh,liu}
based on a smaller width of $D^*$. The coupling constant
$g_{\rho DD}$ is taken to be $g_{\rho DD}=2.52$, which is 
determined in Refs. \cite{matinyan,lin} based on the vector meson
dominance model. The values for both $g_{\pi DD^*}$ and $g_{\rho DD}$
are comparable to those obtained from the QCD sum rules \cite{bracco,navarra}.
 
Other coupling constants, which are not known empirically, are obtained
using SU(4) relations \cite{liu,cscatt}, i.e., 
\begin{eqnarray}
\frac{f_{DN\Lambda_c}}{m_D}&=&\frac{3-2\alpha_D}{\sqrt{3}}
\frac{f_{\pi NN}}{m_{\pi}}\simeq \frac{f_{\pi NN}}{m_{\pi}}=7.18~
{\rm GeV}^{-1},\\
g_{D^{*}N\Lambda_c}&=&\sqrt{3}g_{\rho NN}=5.58,\\
\frac{f_{\pi\Sigma_c\Lambda_c}}{m_{\pi}}&=&\frac{\alpha_D}{\sqrt{3}}
\frac{f_{DN\Lambda_c}}{m_D}\simeq 2.66~{\rm GeV}^{-1},\\
\frac{f_{DN\Sigma_c}}{m_D}&=&(2\alpha_D - 1)\frac{f_{DN\Lambda_c}}{m_D}
=2.01~{\rm GeV}^{-1}.
\end{eqnarray}
where $\alpha_D=D/(D+F)\simeq 0.64$ \cite{adelseck} with $D$ and $F$ 
being the coefficients for the usual $D$-type and $F$-type couplings.

\section{Cross sections}\label{cs}

The amplitudes for the two processes in Fig. \ref{diagram} can be written
as
\begin{eqnarray}
{\cal M}_{1} & = & {\cal M}_{1a}+{\cal M}_{1b}+{\cal M}_{1c},\\
{\cal M}_{2} & = & ({\cal M}^{\mu}_{2a}+{\cal M}^{\mu}_{2b}
+{\cal M}^{\mu}_{2c})\varepsilon_{\mu},
\end{eqnarray}
where $\varepsilon_{\mu}$ is the polarization vector of rho meson.
The amplitudes ${\cal M}_{1a},{\cal M}_{1b}$ and ${\cal M}_{1c}$ 
are for the top three diagrams in Fig. \ref{diagram} and are given by
\begin{eqnarray} 
{\cal M}_{1a} & = & -g_{\pi DD^*}g_{D^{*}N\Lambda_c}(\tau^{i})_{\alpha\beta}
(p_{1}+p_{3})^{\mu}\times\frac{1}{t-m^{2}_{D^*}}[g_{\mu\nu}-
\frac{(p_1-p_3)_{\mu}(p_1-p_3)_{\nu}}{m^{2}_{D^*}}]\nonumber\\
&&\times\bar{\Lambda}_c(p_4)\gamma^{\nu}N(p_2),\\
{\cal M}_{1b} & = & \frac{f_{\pi NN}f_{DN\lambda_c}}{m_{D}m_{\pi}}
(\tau^{i})_{\alpha\beta}\bar{\Lambda}_c(p_4){p\mkern-10mu/}_3
\times\frac{m_N-{q\mkern-10mu/}_s}{s-m^{2}_N}{p\mkern-10mu/}_1N(p_2),\\
{\cal M}_{1c} & = &\frac{f_{\pi\Sigma_c\Lambda_c}f_{DN\Lambda_c}}{m_{\pi}m_D}
(2\delta_{ij}\tau^{j})_{\alpha\beta}\bar{\Lambda}_c{p\mkern-10mu/}_1
\times\frac{m_{\Sigma_c}-{q\mkern-10mu/}_u}{u-m^2_{\lambda_c}}
{p\mkern-10mu/}_{3}N,\\
\end{eqnarray}  
while the amplitudes ${\cal M}^{\mu}_{2a},{\cal M}^\mu_{2b}$, and  
${\cal M}^{\mu}_{2b}$ are for the bottom three diagrams, and they are
\begin{eqnarray}  
{\cal M}^{\mu}_{2a} & = & \frac{-if_{DN\Lambda_c}g_{\rho DD}}
{m_D}(\tau^{i})_{\alpha\beta}(2p_3-p_1)^{\mu}
\times\bar{\Lambda}_c\gamma_{5}\frac{{p\mkern-10mu/}_1-{q\mkern-10mu/}_3}
{t-m^{2}_{D}}N,\\
{\cal M}^{\mu}_{2b} & = & \frac{if_{DN\Lambda_c}g_{\rho NN}}
{m_D}(\tau^{i})_{\alpha\beta}\bar{\Lambda}_c\gamma_5
{p\mkern-10mu/}_3\frac{{q\mkern-10mu/}_s+m_N}{s-m^{2}_{N}}
(\gamma^{\mu}+i\frac{k_{\rho}}{2m_{N}}\sigma^{\nu\mu}p_{1\nu})N(p_2),\\
{\cal M}^{\mu}_{2c} & = & \frac{if_{DN\Sigma_c}g_{\rho\Sigma_c\Lambda_c}}
{m_D}(2\delta_{ij}\tau^{j})_{\alpha\beta}\bar{\Lambda}_c\gamma^{\mu}
\times\frac{{q\mkern-10mu/}_u+m_{\Sigma_c}}{u-m^2_{\Sigma_c}}
\gamma_5{p\mkern-10mu/}_3 N.
\end{eqnarray}  
In the above, $p_1, p_2, p_3$ and $p_4$ denote the momenta of $\pi (\rho)$, 
$N$, $\bar{D}$ and $\Lambda_c$, respectively; $s=(p_1+p_2)^2$, $t=(p_1-p_3)^2$,
and $u=(p_1-p_4)^2$ are the usual Mandelstam variables; and
$q_s=p_1+p_2$ and $q_u=p_2-p_3$.

The isospin- and spin-averaged differential cross sections for the two 
processes in Fig. \ref{diagram} are then 
\begin{eqnarray}
\frac{d\sigma_{\pi N\to \bar{D}\Lambda_c}}{dt} & = &\frac{1}
{768\pi sp^{2}_{i}}|{\cal M}_{1}|^{2},\\
\frac{d\sigma_{\rho N\to \bar{D}\Lambda_c}}{dt} & = &\frac{1}
{2304\pi sp^{2}_{i}}|{\cal M}_{2}|^{2}.
\end{eqnarray}
The squared invariant scattering amplitudes $|{\cal M}_{1}|^{2}$ 
and $|{\cal M}_{2}|^{2}$, which include the summation over the spins and 
isospins of both initial and final particles, can be evaluated 
using the software package $FORM$\cite{form}. In evaluating 
these cross sections, we have introduced form factors at the interaction 
vertices. For three-point vertices, i.e., $\pi DD^*$, $\rho DD$, $\rho NN$, 
$\pi NN$, $DN\Lambda_c$, $D^* N\Lambda_c$, $DN\Sigma_c$, and 
$\rho\Sigma_c\Lambda_c$, they are taken to have the form \cite{cscatt,tsushima}
\begin{eqnarray}
f_1=\frac {\Lambda^2}{\Lambda^2+{\bf q}^2}, ~~~~ f_2=\frac {\Lambda^2}
{\Lambda^2+{\bf p}_i^2}.
\end{eqnarray}
where $f_1$ is for $t$ and $u$ channels and $f_2$ for $s$ channel with 
${\bf q}^2$ and ${\bf p}_i^2$ being, respectively, the squared three 
momentum transfer and squared initial three momentum in the center-of-mass 
frame of the pion or rho meson and nucleon. In studying 
$J/\psi$ absorption and charm meson scattering using the same
interaction Lagrangians \cite{haglin,lin,oh}, values for the cutoff parameter
$\Lambda$ are usually taken to be 1 or 2 GeV. We use these values
in the present study.

\section{results}
\label{results}
  
\begin{figure}[htb]
\centerline{\epsfig{file=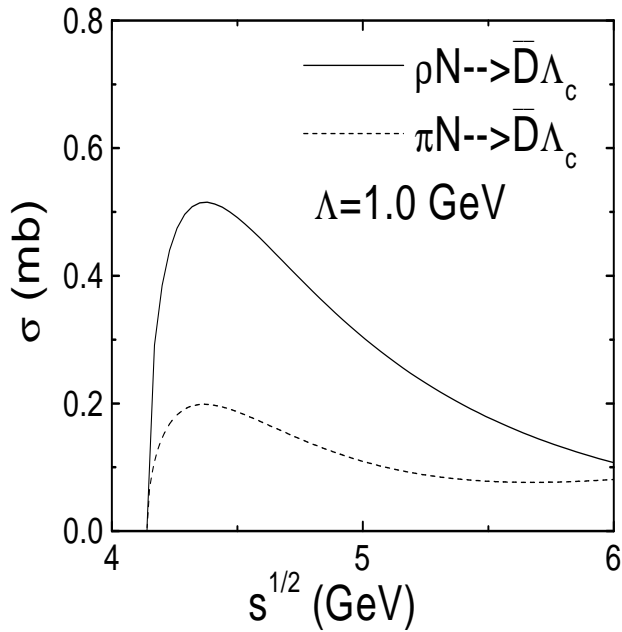,width=3.5in,height=3.5in,angle=-90}}
\vspace{0.5cm}
\caption{Cross sections for charm meson production from meson-nucleon 
scattering as functions of center-of-mass energy for cutoff parameter 
of 1 GeV.}
\label{cross1}
\end{figure}

\begin{figure}[htb]
\centerline{\epsfig{file=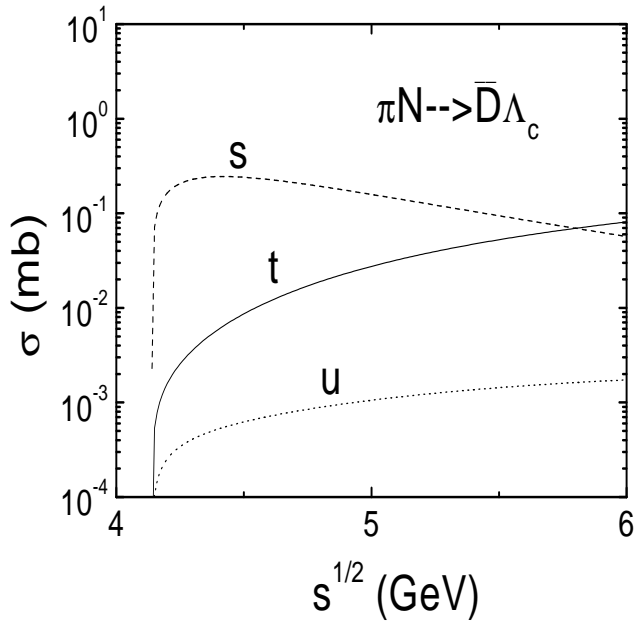,width=3.5in,height=3.5in,angle=-90}}
\vspace{0.5cm}
\caption{Contributions from $t$, $s$, and $u$ channels to the charm meson
production cross section from pion-nucleon scattering as functions of 
center-of-mass energies with cutoff parameter $\Lambda=1$ GeV.}
\label{cross2}
\end{figure}

We first show the results obtained with a cutoff parameter $\Lambda=1$
GeV. In Fig. \ref{cross1}, the cross sections for charm meson production from 
meson-nucleon scattering are given as functions of center-of-mass energy.  
It is seen that the cross section for the reaction $\pi N\to\bar D\Lambda_c$
(dotted curve) has a peak value of about 0.2 mb. Although, this value
is much larger than that predicted by the QGSM model \cite{hcharm}, 
it is mainly due to the $s$ channel that involves a nucleon pole as 
shown by the dashed curve in Fig. \ref{cross2}, where the cross sections 
from individual amplitudes are shown. The contribution from the $t$ channel 
charm vector meson exchange (solid curve) at low center-of-mass energy 
has a similar magnitude as found in QGSM, while the $u$ channel contribution 
(dotted curve) is indeed negligible as assumed in Ref. \cite{hcharm}.

\begin{figure}[htb]
\centerline{\epsfig{file=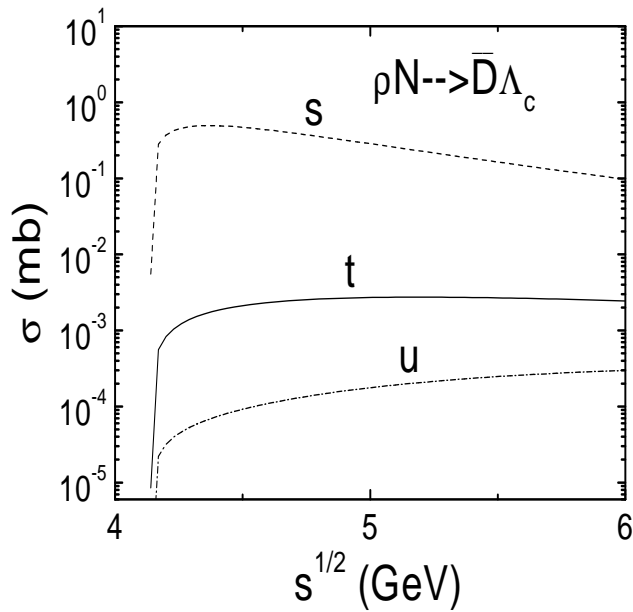,width=3.5in,height=3.5in,angle=-90}}
\vspace{0.5cm}
\caption{Same as Fig. \protect\ref{cross2} charm meson
production cross section from rho-nucleon scattering.}
\label{cross3}
\end{figure}

The cross section for the reaction $\rho N\to\bar D\Lambda_c$
from rho-nucleon scattering shown by the solid curve
in Fig. \ref{cross1} is about a factor of two larger than that from 
pion-nucleon scattering. The relative importance of the contributions
from the $s$, $t$, and $u$ channels in this case is shown in 
Fig. \ref{cross3}. Again, the dominant contribution is from $s$
channel, while the $t$ and $u$ channel contributions are much smaller. 

The magnitude of charm meson production cross sections depends 
strongly on the value of the cutoff parameter. If we use a larger value
of $\Lambda=2$ GeV as suggested by the QCD sum rules \cite{qcd}, these 
cross sections are increased by an order of magnitude. On the other hand, 
their values are reduced by more than an order of magnitude if a smaller 
value of $\Lambda=0.5$ GeV is used. We note that to reproduce the 
empirical cross section for kaon production from pion-nucleon scattering, 
i.e., $\pi N\to K\Lambda$, using the same SU(4) invariant Lagrangian at 
the Born approximation requires $\Lambda\sim 0.4$ GeV. Because of the 
smaller sizes of charm hadrons, we expect, however, that the cutoff 
parameter at interaction vertices involving these particles should 
have a larger value than at those involving strange hadrons. 
Using $\Lambda=1$ GeV for charm meson production thus seems reasonable.

\section{summary}\label{summary}

Using a SU(4) invariant meson-baryon effective Lagrangian, we have
studied the cross section for charm meson production in pion-nucleon
and rho-nucleon scattering. We find that the magnitude of these cross 
sections depends sensitively on the value of the cutoff parameter at 
interaction vertices. With a cutoff parameter of 1 GeV, 
the cross section for $\pi N\to\bar D\Lambda_c$ has a peak value
of about 0.2 mb, while that for $\rho N\to\bar D\Lambda_c$ is
about a factor of two larger. The dominant contribution to these cross 
sections is from the $s$ channel nucleon pole diagram. The contribution 
from the $t$ channel charm meson exchange is less important and has 
a magnitude comparable to that given by the Quark-Gluon 
String Model. The $u$ channel contribution is negligible for charm meson 
production from both pion-nucleon and rho-nucleon scattering. Since our cross 
sections for charm meson production are much larger than that given by 
the QGSM model, they would lead to too large an enhancement of charm 
meson production if used during the initial string stage of heavy ion
collisions as in Ref. \cite{hcharm}. On the other hand, more reasonable 
results for charm production are expected if these cross sections are 
used only for collisions between mesons and baryons in the hadronic matter. 

\section*{acknowledgment}

We thank Wolfgang Cassing for helpful discussions.
This work was supported by the National Science Foundation under Grant 
Nos. PHY-9870038 and PHY-0098805, the Welch Foundation under Grant 
No. A-1358, and the Texas Advanced Research Program under Grant 
No. FY99-010366-0081.

\end{document}